\documentclass[11pt]{article}

\usepackage{amsfonts}
\usepackage[centertags]{amsmath}
\usepackage{amssymb}
\usepackage{cite}
\usepackage{psfrag}
\usepackage{graphicx}
\usepackage{bbm,bm}
\usepackage{epsfig, multicol}
\usepackage{color}
\allowdisplaybreaks
\begin{document}

\topmargin 0pt
\oddsidemargin 0mm
\def\be{\begin{equation}}
\def\ee{\end{equation}}
\def\bea{\begin{eqnarray}}
\def\eea{\end{eqnarray}}
\def\ba{\begin{array}}
\def\ea{\end{array}}
\def\ben{\begin{enumerate}}
\def\een{\end{enumerate}}
\def\nab{\bigtriangledown}
\def\tpi{\tilde\Phi}
\def\nnu{\nonumber}
\newcommand{\eqn}[1]{(\ref{#1})}

\newcommand{\half}{{\frac{1}{2}}}
\newcommand{\vs}[1]{\vspace{#1 mm}}
\newcommand{\dsl}{\pa \kern-0.5em /} 
\def\a{\alpha}
\def\b{\beta}
\def\g{\gamma}\def\G{\Gamma}
\def\d{\delta}\def\D{\Delta}
\def\ep{\epsilon}
\def\et{\eta}
\def\z{\zeta}
\def\t{\theta}\def\T{\Theta}
\def\l{\lambda}\def\L{\Lambda}
\def\m{\mu}
\def\f{\phi}\def\F{\Phi}
\def\n{\nu}
\def\p{\psi}\def\P{\Psi}
\def\r{\rho}
\def\s{\sigma}\def\S{\Sigma}
\def\ta{\tau}
\def\x{\chi}
\def\o{\omega}\def\O{\Omega}
\def\k{\kappa}
\def\pa {\partial}
\def\ov{\over}
\def\nn{\nonumber\\}
\def\ud{\underline}
\def\qq{$Q{\bar Q}$}
\begin{flushright}
%
\end{flushright}
\begin{center}
{\Large{\bf Holographic entanglement entropy and entanglement \\thermodynamics of `black' non-susy D3 brane}}

\vs{10}

{Aranya Bhattacharya\footnote{E-mail: aranya.bhattacharya@saha.ac.in} 
and Shibaji Roy\footnote{E-mail: shibaji.roy@saha.ac.in}}

\vs{4}

{\it Saha Institute of Nuclear Physics\\
1/AF Bidhannagar, Calcutta 700064, India}

\vs{4}

{\rm and}

\vs{4}

{\it Homi Bhabha National Institute\\
Training School Complex, Anushakti Nagar, Mumbai 400085, India}
\end{center}

\vs{10}

\begin{abstract}
Like BPS D3 brane, the non-supersymmetric (non-susy) D3 brane of type IIB string theory is also known to have a decoupling 
limit and leads to a non-supersymmetric AdS/CFT correspondence. The throat geometry in this case represents a QFT which is
neither conformal nor supersymmetric. The `black' version of the non-susy D3 brane in the decoupling limit describes a QFT
at finite temperature. Here we first compute the entanglement entropy for small subsystem of such QFT from the decoupled geometry 
of `black' non-susy D3 brane using holographic technique. Then we study the entanglement thermodynamics for the weakly excited
states of this QFT from the asymptotically AdS geometry of the decoupled `black' non-susy D3 brane. We observe that for
small subsystem this background indeed satisfies a first law like relation with a universal (entanglement) temperature inversely 
proportional to the size of the subsystem and an (entanglement) pressure normal to the entangling surface. Finally we show how
the entanglement entropy makes a cross-over to the thermal entropy at high temperature.                     
\end{abstract}

\newpage

\section{Introduction}

The entanglement entropy (EE) is a measure of quantum information encoded in a quantum system. In particular, for a bipartite
system the EE of a subsystem $A$ is the von Neumann entropy and is defined as $S_A = - {\rm Tr}(\rho_A \log \rho_A)$,
where $\rho_A = {\rm Tr}_B (\rho_{\rm tot})$ is the reduced density matrix on $A$ obtained by tracing out on $B$, the complement of $A$,
of the density matrix of the total system $\rho_{\rm tot}$ (see, for example, \cite{Bombelli:1986rw, Srednicki:1993im, Holzhey:1994we,
Calabrese:2004eu, Eisert:2008ur, Nishioka:2009un, Takayanagi:2012kg} including some reviews). It is useful for many body systems to 
describe various quantum phases
of matter and serves as an order parameter for the quantum phase transitions which occur near zero temperature 
\cite{Vidal, Peschel, Its, Kitaev:2005dm, Levin}. The density matrix
can be carefully defined in the continuum and therefore, EE can be calculated in a QFT in principle using the so-called replica trick 
(see, for example \cite{Rangamani:2016dms}).
However, the actual computation can be done quite generally only in low dimensional CFT$_{d+1}$ ($d < 2$) \cite{Holzhey:1994we, Calabrese:2004eu}. 
For higher dimensions the
computation of EE becomes intractable except for some special cases, like free field QFT in three dimensions and also for CFT$_4$ 
\cite{Rangamani:2016dms}.

Ryu and Takayanagi \cite{Ryu:2006bv, Ryu:2006ef}, motivated by the Bekenstein-Hawking entropy formula, gave a prescription to compute 
EE in any dimensions using
the idea of AdS/CFT \cite{Maldacena:1997re, Aharony:1999ti}. According to them, the holographic EE (HEE) of the subsystem $A$ in 
the gravity dual is given by \cite{Ryu:2006bv}
\be\label{HEE}
S_E = \frac{{\rm Area}(\gamma_A^{\rm min})}{4G_N}
\ee
where $\gamma_A^{\rm min}$ is the $d$-dimensional minimal area (time-sliced) surface in AdS$_{d+2}$ space whose boundary matches with 
the boundary of the subsystem $A$,
i.e., $\partial\gamma_A^{\rm min} = \partial A$ and $G_N$ is the $(d+2)$-dimensional Newton's constant. The HEE given in \eqref{HEE} has 
been checked \cite{Ryu:2006bv} to agree with the 
QFT results in lower dimensions. In higher dimensions also they give correct qualitative behaviors. In thermodynamics the entropy
of a system can be increased by injecting energy to the system, where the proportionality constant is given by the inverse
of temperature. This leads to an energy conservation relation $\Delta E = T \Delta S$, the first law of thermodynamics. An analogous
problem was addressed in \cite{Bhattacharya:2012mi} for the EE, i.e., to see how the EE of a certain region grows with the increase in energy. Here
the EE is computed using AdS/CFT. The excited state of a CFT is given by the deformation of AdS whose EE can be computed using \eqref{HEE}.
This is then compared with the time component of the boundary stress tensor $T_{tt}$ or the energy density. For a small subsystem $A$,
the total energy is found to be proportional to the increase in EE and the proportionality constant is $c/\ell$, where $c$ is a universal 
constant and $\ell$ is the size of the subsystem. This has been identified with the entanglement temperature in analogy with first law of 
thermodynamics \cite{Bhattacharya:2012mi}. However, in \cite{Allahbakhshi:2013rda}, it has been noted that this is not the complete story. 
Since the first law contains more terms
here also $\Delta E$ can have a term analogous to $P\Delta V$ term. Indeed, by calculating the other components of the boundary stress
tensor it has been found that $\Delta E$ contains a term $d/(d+2) V_d \Delta P_x$ for asypmtotically AdS$_{d+2}$ space, where $\Delta P_x$
is the pressure normal to the entangling surface and $V_d$ is the volume. Therefore the analogous entanglement thermodynamical relation
takes the form \cite{Allahbakhshi:2013rda},
\be\label{ethermo}
\Delta E = T_E \Delta S_E + \frac{d}{d+2} V_d \Delta P_x
\ee

In this paper we consider the non-susy D3 brane or, to be precise, a finite temperature version of that solution in type IIB string 
theory \cite{Chakraborty:2017wdh}. It is known that like BPS D3 brane, non-susy D3 brane also has a decoupling limit 
\cite{Nayek:2015tta, Nayek:2016hsi} and therefore, gives a gravity dual of a 
non-supersymmetric finite temperature gauge theory in the decoupling limit. The gauge theory in this case is non-conformal. We use this
gravity dual to compute the EE of the associated QFT from the Ryu-Takayanagi prescription \eqref{HEE}. Since the non-susy D3 brane
in the decoupling limit has an asymptotically AdS$_5$ geometry, the HEE can be written as a pure AdS$_5$ part and additional part
which can be thought of as the EE associated with an excited state. We use Fefferman-Graham coordinate to compute the HEE and this
helps us to identify the boundary stress tensor quite easily \cite{Balasubramanian:1999re, deHaro:2000vlm}. Having identified the 
boundary stress tensor we then check that the
additional EE of the excited state indeed satisfies the first law like thermodynamical relation we just mentioned in \eqref{ethermo}
for small subsystem. We have identified the entanglement temperature in this case which is inversely related to the size of the entangling
region by a universal constant and also an entanglement pressure normal to the entangling surface. Although non-susy D3 brane we are
considering here has a naked singularity, one can define a temperature related to one of the parameters of the solution. When the
parameter takes a particular value the solution reduces to the standard Schwarzschild AdS$_5$ solution and we checked that for that
particular value of the parameter our results reduce to those obtained in earlier works \cite{Bhattacharya:2012mi}. We also checked that 
at higher temperature the HEE makes a cross-over \cite{Swingle:2011mk} to the thermal entropy of standard black D3 brane.

The rest of the paper is organized as follows. In section 2, we briefly discuss the decoupled geometry of `black' non-susy D3 brane solution. 
The Fefferman-Graham coordinate and the computation of HEE is given in section 3. In section 4, we give the boundary stress tensors 
and study the entanglement thermodynamics. The cross-over of the HEE to Bekenstein-Hawking thermal entropy is discussed in section 5. 
Finally we conclude in section 6.
  
\section{Decoupled geometry of `black' non-susy D3 brane}

The `black' non-susy D3 brane solution of type IIB string theory has been discussed in detail in \cite{Chakraborty:2017wdh} and so we will be brief 
here. The purpose 
for our discussion here is to fix the notation and convention for the computation of HEE in the next section. The solution in the Einstein frame
takes the form,
\bea\label{nonsusyd3}
& &  ds^{2}= F_1(\rho)^{-\frac{1}{2}}G(\rho)^{-\frac{\delta_{2}}{8}}\left[-G(\rho)^{\frac{\delta_{2}}{2}}dt^{2}+ \sum_{i=1}^{3}(dx^{i})^{2}\right]+F_1(\rho)^{\frac{1}{2}}
G(\rho)^{\frac{1}{4}}\left[\frac{d\rho^{2}}{G(\rho)}+\rho^{2}d\Omega_{5}^{2}\right]\nn
& & e^{2\phi} = G(\rho)^{-\frac{3\d_2}{2} + \frac{7\d_1}{4}}, \qquad
 \qquad F_{[5]} = \frac{1}{\sqrt{2}}(1+\ast) Q {\rm Vol}(\Omega_5).
\eea
where the functions $G(\rho)$ and $F(\rho)$ are defined as,
\be\label{functions}
G(\rho)=1+\frac{\rho_{0}^{4}}{\rho^{4}},\qquad
F_1(\rho)=G(\rho)^{\frac{\alpha_1}{2}}\cosh^{2}\theta - G(\rho)^{-\frac{\beta_1}{2}}\sinh^{2}\theta 
\ee
Here $\d_1$, $\d_2$, $\a_1$, $\b_1$, $\theta$, $\rho_0$, $Q$ are the parameters characterizing the solution. Now
to compare this solution with that given in eq.(6) of \cite{Chakraborty:2017wdh}, we note that we have replaced $\d$ by $\d_2$ here 
and also, the function $F(\rho)$ there is related to
$F_1(\rho)$ by the relation $F_1(\rho) = G(\rho)^{3\d_1/8} F(\rho)$. The parameters $\a$ and $\b$ there are related to $\a_1$ and $\b_1$ by the relations
$\a_1 = \a + 3\d_1/4$ and $\b_1 = \b - 3\d_1/4$. We point out that the parameters are not all independent but they satisfy the following relations
\bea\label{relations}
& & \a_1 - \b_1 = \a - \b + 3\d_1/2 = 0\nn
& & \a_1 + \b_1 = \a + \b = \sqrt{10 - \frac{21}{2} \d_2^2 - \frac{49}{2} \d_1^2 + 21 \d_2\d_1}\nn
& & Q = (\a_1+\b_1)\rho_0^4 \sinh2\theta
\eea
Note that the solution has a curvature singularity at $\rho=0$ and also the metric does not have the full Poincare symmetry ISO(1, 3) in the brane
world-volume directions, rather, it is broken to R $\times$ ISO(3) and this is the reason we call it `black' non-susy D3 brane solution. However, we put
black in inverted comma because this solution does not have a regular horizon as in ordinary black brane but, has a singular horizon. 
The standard zero temperature non-susy D3 brane solution given in eq.(1) of \cite{Nayek:2016hsi}
can be recovered from \eqref{nonsusyd3} by simply putting $\d_2=0$ and identifying $7\d_1/4$ as $\d$ there. We remark that in spite of 
the solution \eqref{nonsusyd3} has a singular horizon we can still define a temperature as argued in \cite{Kim:2007qk} and by comparing 
the expression for
temperature there we can obtain the temperature of the `black' non-susy D3 brane as,
\be\label{temp}
T_{\rm nonsusy} = \left(\frac{-2\d_2}{(\a_1+\b_1)^2}\right)^{\frac{1}{4}}\frac{1}{\pi \rho_0 \cosh\theta}
\ee
From the above expression it is clear that for the reality of the temperature the parameter $\d_2$ must be less or equal to zero.
It is straightforward to check that when $\d_2=-2$ and $\d_1=-12/7$ (which implies $\a_1 = \b_1 = 1$ and $\a_1+\b_1=2$), the above solution \eqref{nonsusyd3}
reduces precisely to the ordinary black D3 brane solution and the temperature \eqref{temp} also reduces to the Hawking temperature of the ordinary black
D3 brane.                  
   
From now on we will put $\a_1+\b_1 = 2$ for simplicity. Therefore, from the first relation in \eqref{relations}, we have $\a_1=1$ and $\b_1=1$.
In this case, the parameters $\d_1$ and $\d_2$ will be related (see the second equation in \eqref{relations}) by
\be\label{deltareln}
42 \d_2^2 + 49 \d_1^2 - 84 \d_1\d_2 = 24
\ee
The function $F_1(\rho)$ given in \eqref{functions} then reduces to 
\be\label{fonerho}
F_1(\rho) = G(\rho)^{-\frac{1}{2}} H(\rho), \qquad {\rm where,} \qquad H(\rho) = 1 + \frac{\rho_0^4 \cosh^2\theta}{\rho^4} \equiv 1 + \frac{\rho_1^4}{\rho^4}
\ee
Therefore the Einstein frame metric in \eqref{nonsusyd3} reduces to
\be\label{metric1}
ds^{2}= H(\rho)^{-\frac{1}{2}}G(\rho)^{\frac{1}{4}-\frac{\delta_{2}}{8}}\left[-G(\rho)^{\frac{\delta_{2}}{2}}dt^{2}+ \sum_{i=1}^{3}(dx^{i})^{2}\right]+H(\rho)^{\frac{1}{2}}
\left[\frac{d\rho^{2}}{G(\rho)}+\rho^{2}d\Omega_{5}^{2}\right] 
\ee
where $H(\rho)$ is given in \eqref{fonerho}. The decoupled geometry can be obtained by zooming into the region
\be\label{zoom}
\rho \sim \rho_0 \ll \rho_0 \cosh^{\half}\theta
\ee
Note that in this limit $\theta \to \infty$ and the function $H(\rho)$ can be approximated as $H(\rho) \approx \rho_1^4/\rho^4$, but $G(\rho)$
remains unchanged\footnote{Here we remark that since in the decoupling limit \eqref{zoom}, $\theta \to \infty$, we might think from the expression of
temperature in \eqref{temp} that, the temperature goes to zero if all other parameters of the theory, namely, $\a_1,\,\b_1,\,\d_2,\,\rho_0$ are kept fixed.
While this is true, but we must remember that in the decoupling limit $\rho_0$ also goes to zero (see eq.(13) of ref.\cite{Chakraborty:2017wdh}) such 
that their product or the temperature remains finite. However, this finite value can be very large or very small making the temperature of the solution
very small or very large, respectively. So, when we discuss high temperature limit we mean that we set $\rho_0 \cosh\theta$ to a finite but very small
value, keeping other parameters finite and fixed.}. The metric \eqref{metric1} then reduces to,
\be\label{metric2}
ds^{2}= \frac{\rho^2}{\rho_1^2}G(\rho)^{\frac{1}{4}-\frac{\delta_{2}}{8}}\left[-G(\rho)^{\frac{\delta_{2}}{2}}dt^{2}+ \sum_{i=1}^{3}(dx^{i})^{2}\right]+
\frac{\rho_1^2}{\rho^2}\frac{d\rho^{2}}{G(\rho)}+\rho_1^{2}d\Omega_{5}^{2} 
\ee
where $\rho_1 = \rho_0 \cosh^{\half}\theta$ is the radius of the transverse 5-sphere which decouples from the five dimensional asymptotically AdS$_5$
geometry. As the 5-sphere decouples, we will work with the rest of the five dimensional geometry to compute the HEE in the next section.

\section{Holographic entanglement entropy in Fefferman-Graham coordinates}

In this section we first rewrite our asymptotically AdS$_5$ metric (leaving out the 5-sphere part) given by
\be\label{aads5}
ds^{2}= \frac{\rho^2}{\rho_1^2}G(\rho)^{\frac{1}{4}-\frac{\delta_{2}}{8}}\left[-G(\rho)^{\frac{\delta_{2}}{2}}dt^{2}+ \sum_{i=1}^{3}(dx^{i})^{2}\right]+
\frac{\rho_1^2}{\rho^2}\frac{d\rho^{2}}{G(\rho)}
\ee
in the Fefferman-Graham form and then compute the HEE from this geometry. Note that as $\rho \to \infty$, $G(\rho) \to 1$ and the
metric reduces to AdS$_5$ form. The $(d+2)$-dimensional asymptotically AdS space can be written in Fefferman-Graham
coordinates as,
\be\label{FG}
ds^2_{d+2} = \frac{\rho_1^2}{r^2} dr^2 + \frac{r^2}{\rho_1^2} g_{\mu\nu}(x,r) dx^\mu dx^\nu
\ee
where $g_{\mu\nu} = \eta_{\mu\nu} + h_{\mu\nu}(x,r)$ with
\be\label{hmunu}
h_{\mu\nu} (x,r) = h_{\mu\nu}^{(0)} (x) + \frac{1}{r^2} h_{\mu\nu}^{(2)} (x) + \cdots + \frac{1}{r^{d+1}} h_{\mu\nu}^{(d+1)} (x) + \cdots
\ee
and for $d$ = odd, the $(d+3)/2$-th term can contain an additional $\log r$ piece, however, for our solution \eqref{aads5} this
does not appear. Now in order to express \eqref{aads5} in the form of \eqref{FG}, we must change the radial coordinate $\rho$ to
$r$. By inspecting \eqref{aads5} and \eqref{FG} (for $d=3$), we get the relation,
\be\label{rtorho}
\rho^2 + \sqrt{\rho^4 + \rho_0^4} = r^2
\ee
and inverting this relation we get,
\be\label{rhotor}
\rho^2 = \frac{r^2}{2} - \frac{\rho_0^4}{2r^2}
\ee
By scaling $r \to \sqrt{2} r$, the above relation reduces to
\be\label{scaledr}
\rho^2 = r^2\left(1-\frac{\rho_0^4}{4 r^4}\right)
\ee
Using \eqref{scaledr} the metric \eqref{aads5} takes the form,
\be\label{metric3}
ds^{2}=\frac{r^{2}}{\rho_{1}^{2}}\left[-\left(1+\frac{3\delta_{2}}{8}\frac{\rho_{0}^{4}}{r^{4}}\right)dt^{2}+\left(1-\frac{\delta_{2}}{8}
\frac{\rho_{0}^{4}}{r^{4}}\right)\sum_{i=1}^{3}
(dx^{i})^{2}\right]+\frac{\rho_{1}^{2}}{r^{2}}dr^{2}
\ee  
Since here we are considering only weakly excited states, our geometry will be near the boundary and that is the reason as a first
order approximation we have replaced $(1-\rho_0^4/r^4)^a$ by $(1-a\rho_0^4/r^4)$ for any real number $a$ in writing the metric \eqref{metric3}. 
This choice is also needed so that we can apply Ryu-Takayanagi prescription for the calculation of EE \cite{Ryu:2006ef}. To compute HEE, we 
choose another
coordinate $z$ by the relation $z=\rho_1^2/r$ and rewrite the metric in the following form,
\be\label{metric4}
ds^{2}=\frac{\rho_1^{2}}{z^2}\left[-\left(1+\frac{3\delta_{2}}{8}\frac{z^4}{z_0^4}\right)dt^{2}+\left(1-\frac{\delta_{2}}{8}
\frac{z^4}{z_0^4}\right)\sum_{i=1}^{3}
(dx^{i})^{2} + dz^{2}\right]
\ee 
where $z_0^4 = \rho_1^8/\rho_0^4$. This is the form of the metric in Fefferman-Graham coordinates.

Now to compute the holographic EE, we have to first calculate the minimal area of the surface embedded in the time slice
of the background \eqref{metric4} bounded by the edge of $A$, i.e., $\partial A$ which is a strip given by $-\ell/2 \leq
x_1 \leq \ell/2$ and $0 \leq x_{2,3} \leq L$. We parameterize the surface $\gamma_A$ by $x_1=x_1(z)$, then the area of the
embedded surface would be given as,
\be\label{area1}
{\rm Area}(\gamma_A) = \int dx_1\,dx_2\,dx_3 \sqrt{g}
\ee
where $g$ is the determinant of the metric induced on $\gamma_A$. For the strip and for the parameterization mentioned above, the area reduces to
\be\label{area2}
{\rm Area}(\gamma_A) = \rho_1^3 \int dx_2\, dx_3\, dz \frac{\sqrt{\left[1+\left(1-\frac{\delta_2}{8}\frac{z^4}{z_0^4}\right)x_1^{\prime 2}\right]
\left(1-\frac{\delta_{2}}{4}\frac{z^{4}}{z_{0}^{4}}\right)}}{z^3}  
\ee
Here `prime' denotes the derivative with respect to $z$.
Now since $x_1$ is a cyclic coordinate in the above integral \eqref{area2}, we get a constant of motion as follows,
\be\label{const}
 \left(\frac{\rho_1}{z}\right)^3\frac{\left(1-\frac{\delta_2}{4}\frac{z^4}{z_0^4}\right)x_1^{\prime}}{\sqrt{\left[1+\left(1-\frac{\delta_{2}}{8}
\frac{z^{4}}{z_{0}^{4}}\right)x_{1}^{\prime 2}\right]}}=k={\rm constant}
 \ee
Solving this we get $x_{1}^{\prime}$ to be
 \be\label{soln}
  x_{1}^{\prime}= \frac{k}{\sqrt{\left(1-\frac{\delta_{2}}{8}\frac{z^{4}}{z_{0}^{4}}\right)\left[\left(1-\frac{3\delta_{2}}{8}\frac{z^{4}}{z_{0}^{4}}\right)
\left(\frac{\rho_{1}}{z}\right)^{6}-k^{2}\right]}}
\ee    
Actually here we are considering the hanging string configuration given by \eqref{soln} in which the two end points of the string is at the
boundary $z=0$ and has a turning point at $z_\ast$ where $dz/dx$ vanishes. This determines the constant of motion $k$ in terms of $z_\ast$ as,
\be\label{kz}
   k^{2}= \left(1-\frac{3\delta_{2}}{8}\frac{z_{*}^{4}}{z_{0}^{4}}\right)\left(\frac{\rho_{1}}{z_{*}}\right)^{6}
  \ee
Substituting this value of $k$ in \eqref{soln} and integrating we find the size of the entangling region in terms of $z_\ast$ as,
\be\label{lzast}
\ell = 2\int_0^{z_\ast} dz \frac{\left(1-\frac{3\d_2}{16}\frac{z_\ast^4}{z_0^4}\right)}{\sqrt{\left(1-\frac{\d_2}{8}\frac{z^4}{z_0^4}\right)
\left[\frac{z_\ast^6}{z^6} - 1 -\frac{3\d_2}{8}\frac{z_\ast^4}{z_0^4}\left(\frac{z_\ast^2}{z^2} - 1\right)\right]}}
\ee
We will assume that the subsystem is very small such that $\ell \ll z_0$ which amounts to the condition that $\gamma_A$ is close to
the asymptotically AdS$_5$ boundary. We note from the above that when the parameter $\d_2$ related to the temperature of the non-susy
D3 brane (see eq.\eqref{temp}) is put to zero, the metric in \eqref{metric4} reduces to that of AdS$_5$ and the constant of motion
\eqref{kz}, i.e., the turning point $z_\ast$ as well as the size of the entangling region in terms of $z_\ast$ \eqref{lzast}, take the same
forms as those of AdS$_5$ case. Therefore, $\d_2 \neq 0$ solutions are the deformations of AdS$_5$ and represent excited states in the
boundary theory. The above relation \eqref{lzast} can be simplified (as $z,\,z_\ast \ll z_0$) as,
\bea\label{ell}
\ell &=& 2 \int_0^{z_\ast} dz \frac{z^3/z_\ast^3}{\sqrt{1-\frac{z^6}{z_\ast^6}}}\left[1-\frac{3\d_2}{16}\frac{z_\ast^4}{z_0^4}
+ \frac{\d_2}{16}\frac{z^4}{z_0^4} + \frac{3\d_2}{16}\frac{z^4}{z_0^4}\frac{1}{\left(1+\frac{z^2}{z_\ast^2}+\frac{z^4}{z_\ast^4}\right)}
+ \cdots\right]\nn
&=& \frac{2\sqrt{\pi} \Gamma\left(\frac{2}{3}\right)}{\Gamma\left(\frac{1}{6}\right)} z_{\ast(AdS_5)} + \delta z_\ast
\eea
where $z_{\ast(AdS_5)}$ is the turning point for AdS$_5$ and $\delta z_{\ast}$ is the deviation from that value. It has been shown earlier
that there is no change of EE upto the first order due to this change of the turning point from AdS$_5$. So, in evaluating EE we will
use the turning point corresponding to AdS$_5$ only and omit the subscript `AdS$_5$' for brevity. To compute EE, we use the value
of $k$ from \eqref{kz} in \eqref{soln} and substitute it in \eqref{area2} to first obtain the minimized area as,
\be\label{area3}
{\rm Area}(\gamma_A^{\rm min})=2 \int_0^{L} dx_{2} \int_0^{L} dx_{3} \int_{\epsilon}^{z_\ast} dz \left(\frac{\rho_{1}}{z}\right)^{6}
\sqrt{\frac{\left(1-\frac{5\delta_{2}}{8}
\frac{z^{4}}{z_{0}^{4}}\right)}{\left(1-\frac{3\delta_{2}}{8}\frac{z^{4}}{z_{0}^{4}}\right)\left(\frac{\rho_{1}}{z}\right)^{6}-
\left(1-\frac{3\delta_{2}}{8}
\frac{z_\ast^{4}}{z_{0}^{4}}\right)\left(\frac{\rho_{1}}{z_\ast}\right)^{6}}}
\ee       
where $\epsilon$ is an IR cut-off and then use \eqref{HEE} to obtain the EE upto first order in $z^4/z_0^4$ as,
\be\label{HEEf}
   S_{E}=S_{E(0)}+\frac{\rho_{1}^{3}L^{2}}{4G_{(5)}} \int_{0}^{z_\ast} dz \left[\frac{\frac{(-3\d_{2})z^{4}}{8z_{0}^{4}}}{z^{3}\sqrt{1-\frac{z^{6}}{z_\ast^{6}}}}
+\frac{\frac{\d_{2}z^{4}}{8z_{0}^{4}}\sqrt{1-\frac{z^{6}}{z_\ast^{6}}}}{z^{3}}\right]
  \ee
In the above 
\be\label{HEEads5}
 S_{E(0)}=\frac{2\rho_{1}^{3}L^{2}}{4G_{(5)}}\int_{\epsilon}^{z_\ast}\frac{dz}{z^{3}\sqrt{1-(\frac{z}{z_\ast})^{6}}}
\ee
is the EE of the pure AdS$_5$ background. Note that $S_{E(0)}$ is divergent and that is the reason we put an IR cutoff at $\epsilon$ to make it
finite, but the additional term in \eqref{HEEf} is divergence free and we can evaluate the integrals to get the change in EE as,
\bea\label{deltaE}
\Delta S_{E} &=& \frac{\rho_{1}^{3}L^{2}}{4G_{(5)}} z_{\ast}^{2} \left[\frac{(-3\d_{2})\sqrt{\pi}\Gamma\left(\frac{1}{3}\right)}
{48z_{0}^{4}\Gamma\left(\frac{5}{6}\right)}+\frac{\d_{2}\sqrt{\pi}\Gamma\left(\frac{1}{3}\right)}{80z_{0}^{4}\Gamma\left(\frac{5}{6}\right)}\right]\nn
&=& \frac{\rho_{1}^{3}L^{2}}{4G_{(5)}}z_{\ast}^{2}\frac{(-\d_{2})\sqrt{\pi}\Gamma\left(\frac{1}{3}\right)}{20z_{0}^{4}\Gamma\left(\frac{5}{6}\right)}
\eea
Here $z_\ast$ is the value of the turning point for pure AdS$_5$ background given by 
\be\label{zastads5}
   z_{\ast}=\frac{\ell \Gamma\left(\frac{1}{6}\right)}{2\sqrt{\pi}\Gamma\left(\frac{2}{3}\right)}
  \ee
Using this in \eqref{deltaE} we get,
\be\label{deltaE1}
\Delta S_{E}= \frac{(-\d_2)\rho_{1}^{3}L^{2} \ell^2}{320 \sqrt{\pi}G_{(5)} z_0^4}\frac{\Gamma^{2}\left(\frac{1}{6}\right) \Gamma\left(\frac{1}{3}\right)}
{\Gamma^{2}\left(\frac{2}{3}\right)\Gamma\left(\frac{5}{6}\right)}
  \ee
This is the change in the EE of the decoupled theory associated with the `black' non-susy D3 brane from the pure AdS$_5$ solution. We remark that
as $\d_2 = 0$ implies from \eqref{temp} that the temperature of the non-susy D3-brane vanishes, $\Delta S_E$ given in \eqref{deltaE1} also vanishes. This
means that the zero temperature non-susy D3 brane also has vanishing $\Delta S_E$, similar to the case of ordinary black D3 brane, where it
vanishes when the temperature goes to zero.
  
As we have already mentioned in section 2, the non-susy D3 brane solution can be reduced to standard black D3 brane solution when the parameters 
take the values $\d_2=-2$ and $\d_1= -12/7$. Simply taking this limit in \eqref{deltaE1}, we find that the change in EE takes the form,
\be\label{deltaE1black}
\Delta S_{E}= \frac{\rho_{1}^{3}L^{2} \ell^2}{160 \sqrt{\pi}G_{(5)} z_0^4}\frac{\Gamma^{2}\left(\frac{1}{6}\right) \Gamma\left(\frac{1}{3}\right)}
{\Gamma^{2}\left(\frac{2}{3}\right)\Gamma\left(\frac{5}{6}\right)}
  \ee 
This result can be compared with that given in \cite{Bhattacharya:2012mi} and we find that they indeed match once we identify $1/z_0^4 = m$ and $d=4$. 

\section{Entanglement thermodynamics}

As we mentioned in section 3, the asymptotically AdS space in $(d+2)$ dimensions can be expressed in Fefferman-Graham coodinates and it
is given in \eqref{FG}. In this coordinate one can extract the form of boundary stress tensor as follows \cite{Balasubramanian:1999re, deHaro:2000vlm},
\be\label{stensor}
\langle T_{\mu\nu}^{(d+1)} \rangle = \frac{(d+1) \rho_1^d}{16\pi G_{(d+2)}} h_{\mu\nu}^{(d+1)}
\ee
The decoupled `black' non-susy D3 brane geometry (leaving out the S$^5$ part) in Fefferman-Graham coordinate is given in \eqref{metric4}.
So, using this general formula \eqref{stensor} for \eqref{metric4}, we can write down the form of the stress tensor for the boundary theory of 
`black' non-susy D3 brane as,
\be\label{ourstensor}
\langle T_{tt}\rangle = \frac{-3 \rho_1^3 \d_2}{32\pi G_{(5)}}, \qquad \langle T_{x_i x_j}\rangle = \frac{- \rho_1^3 \d_2}{32\pi G_{(5)}} \delta_{ij}, \quad
{\rm where} \quad i,\,j=1,\,2,\,3.
\ee
As we mentioned before, since the parameter $\d_2 \leq 0$, both temporal as well as spatial components of the stress tensor are positive semi-definite.  
Also since here we are considering AdS$_5$, we have put $d=3$ in \eqref{stensor}. Now using these values \eqref{ourstensor} we can rewrite the change in EE 
given by the first expression in \eqref{deltaE} as,
 \be\label{deltase1}
   \Delta S_{E}=\frac{L^{2}z_{\ast}^{2}\pi^{\frac{3}{2}}}{6}\frac{\Gamma\left(\frac{1}{3}\right)}{\Gamma\left(\frac{5}{6}\right)}
\left[\langle T_{tt}\rangle-\frac{3}{5}\langle T_{x_{1}x_{1}}\rangle\right]
  \ee
Putting the value of $z_\ast$ from \eqref{zastads5} we get, 
\be\label{deltase2}
\Delta S_{E}=\frac{L^{2}\ell^{2} \sqrt{\pi}}{24}\frac{\Gamma^{2}\left(\frac{1}{6}\right)\Gamma\left(\frac{1}{3}\right)}{\Gamma^{2}\left(\frac{2}{3}\right)
\Gamma\left(\frac{5}{6}\right)}\left[\langle T_{tt}\rangle -\frac{3}{5}\langle T_{x_{1}x_{1}}\rangle \right]
\ee
In terms of the stress tensors the change in total energy and the pressure are defined as,
\be\label{defn}
\Delta E = L^2 \ell \langle T_{tt}\rangle, \qquad \Delta P_{x_1 x_1} = \langle T_{x_1 x_1} \rangle
\ee
Using this in \eqref{deltase2} we get the change in EE as,
\be\label{deltase3}
\Delta S_{E}=\ell\frac{\sqrt{\pi}}{24}\frac{\Gamma^{2}\left(\frac{1}{6}\right)\Gamma\left(\frac{1}{3}\right)}{\Gamma^{2}\left(\frac{2}{3}\right)
\Gamma\left(\frac{5}{6}\right)}\left[\Delta E - \frac{3}{5} \Delta P_{x_1 x_1} V_3\right]
\ee
where $V_3 = L^2 \ell$ is the volume of the subspace.
Comparing this with the first law of thermodynamics we identify the entanglement temperature to be
\be\label{etemp}
T_{E}=\frac{24 \Gamma\left(\frac{5}{6}\right)\Gamma^{2}\left(\frac{2}{3}\right)}{\ell\sqrt{\pi}\Gamma\left(\frac{1}{3}\right)
\Gamma^{2}\left(\frac{1}{6}\right)}
\ee
We note that the entanglement temperature is inversely proportional to the size $\ell$ of the entangling region with a universal 
proportionality constant \cite{Bhattacharya:2012mi}.
Thus from here, we conclude that the decoupled theory of `black' non-susy D3 brane satisfies the first law of entanglement thermodynamics
\be\label{ethermodynamics}
 \Delta E = T_{E}\Delta S_{E} + \frac{3}{5} \Delta P_{x_{1}x_{1}} V_{3}
\ee
This is indeed the relation we mentioned in \eqref{ethermo} for $d=3$ \cite{Allahbakhshi:2013rda}. 

\section{Cross-over to thermal entropy}
 
In this section we will show how the total HEE of the decoupled theory of `black' non-susy D3 brane we calculated in 
\eqref{HEEf} reduces to thermal entropy of that of standard black D3 brane in the high temperature limit. For this purpose we first look at the
expression for the size of the entangling region in \eqref{lzast}. By defining $z/z_\ast$ as $x$, the integral can be written as,
\bea\label{lzast1}
\frac{\ell}{2} &=& z_\ast \int_0^1 dx \frac{x^3\left(1-\frac{3\d_2}{16}\frac{z_\ast^4}{z_0^4}\right)}{\sqrt{\left(1-\frac{\d_2}{8}\frac{z_\ast^4}{z_0^4} x^4\right)
\left[1 - x^6 -\frac{3\d_2}{8}\frac{z_\ast^4}{z_0^4} x^4\left(1-x^2\right)\right]}}\nn
&=& z_\ast {\cal I}\left(\frac{z_\ast}{z_0}\right)  
\eea
On the other hand the total area integral given in \eqref{area3} can be written as, 
\bea\label{area4}
{\rm Area}(\gamma_A^{\rm min})&=&\frac{2\rho_1^3 L^2}{z_\ast^2} \int_0^1 dx \frac{1}{x^3} 
\sqrt{\frac{\left(1-\frac{5\delta_{2}}{8}
\frac{z_\ast^{4}}{z_{0}^{4}} x^4\right)}{\left(1-\frac{3\delta_{2}}{8}\frac{z_\ast^{4}}{z_{0}^{4}} x^4\right)-
\left(1-\frac{3\delta_{2}}{8}
\frac{z_\ast^{4}}{z_{0}^{4}}\right)x^6}}\nn
&=& \frac{2\rho_1^3 L^2}{z_\ast^2} {\cal {\tilde I}}\left(\frac{z_\ast}{z_0}\right)
\eea
Let us first clarify what we mean by low temperature and high temperature limit. We have seen that 
the temperature of the solution as mentioned earlier (see \eqref{temp}) behaves like $T_{\rm nonsusy}\sim \frac{1}{\rho_0\cosh\theta} = \frac{1}{z_{0}}$ 
and $z_\ast\sim \ell$. Therefore, the ratio $\frac{z_\ast}{z_0}$ is $\sim \ell T_{\rm nonsusy}$. As we mentioned in footnote 3, the product $\rho_0\cosh\theta = z_0$ 
remains finite in the decoupling limit, but
this finite value could be large or small producing a small or large temperature. Thus in the low temperature regime $z_0$ is large
and this corresponds to taking $\frac{z_\ast}{z_{0}}\sim \ell T_{\rm nonsusy} \ll 1$. On the other hand, in the high temperature regime $z_0$ is small 
and this corresponds to $\frac{z_\ast}{z_{0}}\sim \ell T_{\rm nonsusy} \to 1$. Note that the last condition implies $T_{\rm nonsusy} \sim \frac{1}{\ell}$ and so for
small subsystem the temperature is indeed very large. Now it can be seen that in
the high temperature limit ($\frac{z_\ast}{ z_0} \to 1$) both the integrals ${\cal I}$ and ${\cal {\tilde I}}$ are dominated by the pole at $x=1$
and therefore have the same values, i.e., in this limit
\be\label{itildei}
{\cal I}\left(\frac{z_\ast}{z_0}\right) \approx {\cal {\tilde I}}\left(\frac{z_\ast}{z_0}\right)
\ee
From the second expression of \eqref{lzast1} we, therefore, have
\be\label{i}
{\cal I}\left(\frac{z_\ast}{z_0}\right) = \frac{\ell}{2z_\ast} \approx {\cal {\tilde I}}\left(\frac{z_\ast}{z_0}\right)
\ee
Using this in the second expression of \eqref{area4} and then dividing by $4G_{(5)}$, we get the EE at high temperature as,
\be\label{highs}
S_E = \frac{{\rm Area}(\gamma_A^{\rm min})}{4 G_{(5)}} = \frac{\rho_1^3 L^2 \ell}{4G_{(5)} z_\ast^3} = \frac{\pi^3\rho_1^3 V_3 }{4G_{(5)} (\pi z_0)^3}
\ee
Now using the five dimensional Newton's constant $G_{(5)} = (\pi \rho_1^3)/(2 N^2)$, where $N$ is the number of branes and $1/(\pi z_0) = T$,
where $T$, is the temperature of the standard black D3 brane, we get from \eqref{highs}
\be\label{thermals}
\frac{S_E}{V_3} = \frac{\pi^2}{2} N^2 T^3
\ee
the thermal entropy of a standard black D3 brane. This clearly shows that at high temperature the entanglement entropy of a non-susy D3 brane
makes a cross-over \cite{Swingle:2011mk, Dong:2012se} to the thermal entropy of a black D3 brane\footnote{It should be noted that to recover 
thermal entropy from the entanglement entropy the crucial step we use is to take the limit $\frac{z_\ast}{z_0} \sim \ell T_{\rm nonsusy} \to 1$. The entanglement
entropy and thermodynamics is studied (in sections 3 and 4) in the regime $\frac{z_\ast}{z_0} \sim \ell T_{\rm nonsusy} \ll 1$ and one can go to the
former regime in two different ways, namely, either taking the high temperature limit (as we have done in this section) or taking the large
subsystem limit, i.e., taking $\ell$ large. In the latter case, the minimal surface probes deeper into the bulk picking up thermal contribution as
is the case for black D3 brane with regular horizon.}. 

\section{Conclusion}

To conclude, in this paper we have computed the entanglement entropy of a QFT obtained from the decoupled geometry of `black' non-susy D3
brane using holographic prescription of Ryu and Takayanagi. The field theory in this case is non-supersymmetric and non-conformal and we
have considered only a strip-like subsystem. We have used Fefferman-Graham coordinate to compute the entanglement entropy. For a small 
subsystem we have shown that the total EE can be split into a pure AdS$_5$ part and an additional part corresponding to weakly excited
state of the field theory. The additional part was then found to match exactly with the earlier result for ordinary black D3 brane when
the parameter $\d_2$ of the non-susy D3 brane takes a value $-2$. In the Fefferman-Graham coordinate we have obtained the forms of the boundary
stress tensor of the non-susy D3 brane. Using the expressions of the stress tensor and identifying various components with the energy
and pressure densities we have shown that the EE of the excited state satisfies the first law of entanglement thermodynamics proposed earlier.
We have also checked that at high temperature the total EE of the decoupled theory of non-susy D3 brane reduces to the thermal entropy
of that of the ordinary black D3 brane and not the `black' non-susy D3 brane. It is interesting to note that at high temperature the EE of
a non-susy D3 brane prefers to cross-over to the thermal entropy of the ordinary black D3 brane, among all possible non-supersymmetric D3 brane
configurations (with different values of $\d_2$).               

\noindent\section*{Acknowledgements}

One of us (AB) would like to thank Rohit Mishra for useful discussions. We would like to thank the anonymous referee for the useful comments 
which has helped us, we hope, to improve the manuscript.

\end{document}